\documentclass[twocolumn,english,twocolumn,english,aps,prb,floats,showpacs]{revtex4}
\usepackage[T1]{fontenc}
\usepackage[latin9]{inputenc}
\usepackage{amssymb}
\usepackage{esint}
\usepackage{graphicx}
\usepackage{bm}
\makeatletter
\@ifundefined{textcolor}{}
{%
 \definecolor{BLACK}{gray}{0}
 \definecolor{WHITE}{gray}{1}
 \definecolor{RED}{rgb}{1,0,0}
 \definecolor{GREEN}{rgb}{0,1,0}
 \definecolor{BLUE}{rgb}{0,0,1}
 \definecolor{CYAN}{cmyk}{1,0,0,0}
 \definecolor{MAGENTA}{cmyk}{0,1,0,0}
 \definecolor{YELLOW}{cmyk}{0,0,1,0}
}

\makeatother

\usepackage{babel}
\begin{document}

\title{Quantum Tunneling of the Magnetic Moment in the S/F/S Josephson ${\bm \varphi}_{\bf 0}$-Junction}

\author{Eugene M. Chudnovsky}

\affiliation{Physics Department, Herbert H. Lehman College and Graduate School, The City University of New York \\
 250 Bedford Park Boulevard West, Bronx, New York 10468-1589, USA}

\date{\today}
\begin{abstract}
We show that the S/F/S Josephson $\varphi_0$-junction permits detection of macroscopic quantum tunneling and quantum oscillation of the magnetic moment by measuring the ac voltage across the junction. Exact expression for the tunnel splitting renormalized by the interaction with the superconducting order parameter is obtained. It is demonstrated that magnetic tunneling may become frozen at a sufficiently large $\varphi_0$. The quality factor of quantum oscillations of the magnetic moment due to finite Ohmic resistance of the junction is computed. It is shown that magnetic tunneling rate in the $\varphi_0$-junction can be controled by the supercurrent, with no need for the magnetic field. 
\end{abstract}


\maketitle

Quantum tunneling of the magnetic moment has been subject of intensive research due to the fundamental interest in the phenomenon and because of its potential applications for quantum information technology. Early work focused on non-thermal magnetic relaxation in nanoparticles \cite{MQT-book}. Later on, the focus switched to molecular magnets \cite{MolMag-book}. Due to identical structure of the building blocks of crystals made of magnetic molecules, they permit macroscopic studies of quantum tunneling and quantum oscillations of molecular magnetic moments. On the contrary, the reliable observation and quantitative analysis of the quantum tunneling of the magnetic moment in nanoparticles is hampered by the practical impossibility to make identical nanoparticles. Best samples available still have distribution of sizes and other parameters of the particles within at least 20\%.  Due to the exponential dependence of the tunneling rate on the size, this translates into the distribution of tunneling rates within many orders of magnitude. 

Early on, the difficulty mentioned above prompted researchers to look into the possibility to observe magnetic tunneling in individual nanopartices. Such measurements of $10$-nm ferrimagnetic partices of total spin $S \sim 10^5$, deposited on a nanobridge of a dc-SQUID, were pioneered by Wernsdorfer et al. \cite{Wern-PRL97}. The energy barrier was controlled by the external magnetic field. At very small barriers the evidence of non-thermal magnetic relaxation below 1K was obtained. After a preliminary success, however, these efforts were largely abandoned in favor of detecting spin tunneling in better characterized individual magnetic molecules.  Measurements of transport current through magnetic molecules bridged between conducting leads \cite{vinklyrub12nat} and through molecules grafted on carbon nanotubes \cite{ganklyrub13NatNano} produced convincing evidence of the effect. Observation and control of quantum tunneling of the magnetic moment in nanoscale magnetic clusters, however, remains a challenging experimental task.

In S/F/S nanostructures ferromagnetism can affect superconductivity via the magnetic field it generates \cite{Houzet-PRL08,Petkovic-PRB09,CC-PRB10}. Somewhat stronger influence of ferromagnetism on superconductivity at the F/S interface may occur due to the proximity effect \cite{Buzdin2005}. The opposite influence of the superconductivity on the ferromagnetism is typically weak. This can be understood from the fact that the exchange interaction responsible for magnetic ordering is typically of the order of hundreds or thousands of kelvin while interactions responsible for conventional superconductivity are in the ball park of a few kelvin. One should notice, however, that relativistic interactions responsible for the orientation of the magnetic moment in ferromagnets -- the magnetic anisotropy energy is also in the kelvin, or can be even in the subkelvin, range. Thus, the coupling of the superconducting order parameter to the orientation of the magnetic moment can, in principle, produce a formidable torque on the moment. 

Such a possibility was recently proposed by Buzdin who noticed that spin-orbit interaction in a ferromagnet without inversion symmetry provides the coupling between the direction of the magnetic moment and the superconducting order parameter \cite{Buzdin-PRL08,Buzdin-PRL09}. In a non-centrosymmetric ferromagnetic junction, that Buzdin called the $\varphi_0$-junction, the time reversal symmetry is broken, and the current-phase relation becomes $I = I_c\sin(\varphi - \tilde{\varphi}_0)$, where the phase shift $\tilde{\varphi}_0$ is proportional to the component of the magnetic moment perpendicular to the gradient of the asymmetric spin-orbit potential. In this Letter we argue that such Josephson junction is ideally suited for the study of quantum tunneling of the magnetic moment. The magnetic tunneling would show in the ac voltage across the junction and it can be controlled by the dc supercurrent through the junction. 

\begin{figure}[htbp!]
\center
\includegraphics[scale=0.35]{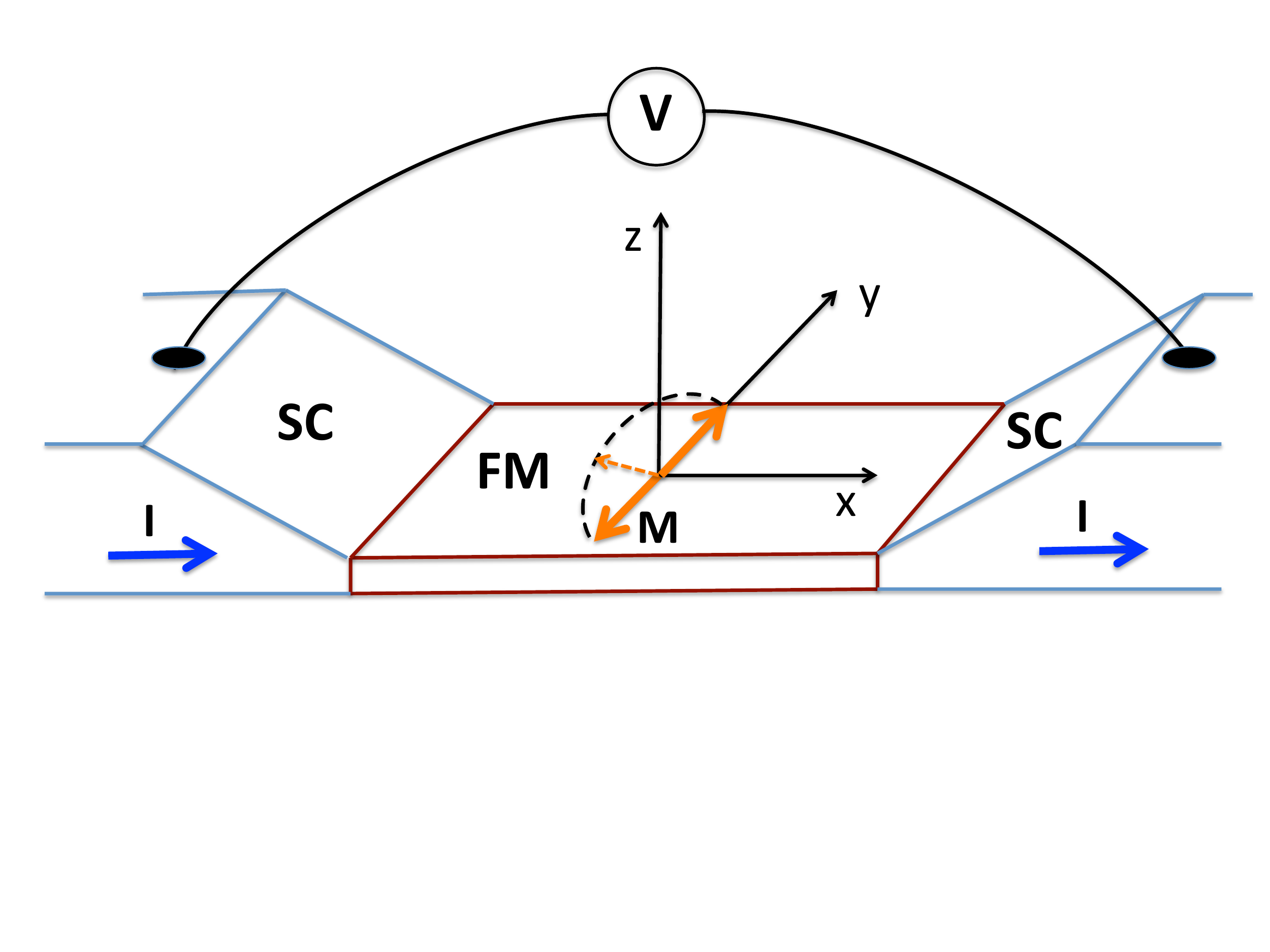}
\vspace{-2cm}
\caption{Schematic picture of the Josephson $\varphi_0$-junction.}
\label{geometry}
\end{figure}
Following Buzdin \cite{Buzdin-PRL09} we consider a S/F/S Josephson $\varphi_0$-junction depicted in Fig.\ \ref{geometry},  
with the potential energy
\begin{equation}\label{U}
U = E_J\left\{1-\cos[\varphi - \tilde{\varphi}_0({\bf M})] - \frac{I}{I_c}\varphi\right\} + U_M({\bf M}) 
\end{equation}
Here $U_M$ is the magnetic anisotropy energy with ${\bf M}$ being the magnetization of the ferromagnetic layer, $\tilde{\varphi}_0$ depends on the direction of ${\bf M}$, and $E_J = I_c\Phi_0/(2\pi)$ is the Josephson energy, with $\Phi_0$ being the flux quantum. With the gradient of the spin-orbit potential normal to the layer (that is, along the $Z$-direction), $\tilde{\varphi}_0$ is given by \cite{Buzdin-PRL09}
\begin{equation}
\tilde{\varphi}_0 = \varphi_0\left(\frac{M_y}{M_0}\right), \qquad \varphi_0 \equiv   l\left(\frac{v_{so}}{v_F}\right),
\end{equation}
where $M_0$ is the length of the magnetization, $v_{so}/v_F \ll 1$ characterizes the strength of the spin-orbit interaction, and $l = 4LE_{ex}/(\hbar v_F)$, with $L$ being the length of the ferromagnetic layer in the $X$-direction and $E_{ex}$ being the energy of the exchange interaction between conducting electrons and localized ferromagnetic spins.  Typically  $v_{so}/v_F \sim 0.1$, $L \lesssim 10$nm, and $E_{ex} \sim 100-500$K, so $\varphi_0$ should be in the range $0.01 < \varphi_0 < 0.1$ \cite{Buzdin-PRL09}.

In this Letter we are making three main points.  The first is that the interaction between the magnetic moment and the superconducting order parameter in the $\varphi_0$-junction renormalizes the tunnel splitting in a manner that can be exactly computed and measured. The second point is that the Josephson $\varphi_0$-junction permits detection of the quantum tunneling and quantum oscillations of the magnetic moment by measuring the voltage across the junction. The third point is that the $\varphi_0$-junction allows one to control the magnetic tunneling rate by the supercurrent through the junction. To illustrate the first two points we chose a typical form of $U_M$ for a ferromagnetic layer that corresponds to the $XY$ easy magnetization plane with the $Y$ easy axis in that plane,
\begin{equation}
U_M = \frac{1}{2}K_{\perp}V\left(\frac{M_z}{M_0}\right)^2 - \frac{1}{2}K_{\parallel}V\left(\frac{M_y}{M_0}\right)^2,
\end{equation}
$V$ being the volume of the ferromagnetic layer. At $I = 0$ classical degenerate equilibrium corresponds to two opposite orientations ${\bf M}$ along the $Y$-axis, with the energy barrier betwen them given by $U_0 = \frac{1}{2}K_{\parallel}V$.  The current may reduce the barrier to zero and, thus, it can switch the direction of ${\bf M}$ \cite{Buzdin-PRL08,Buzdin-PRL09}. Here we are interested in the quantum switching of ${\bf M}$ at a finite energy barrier. 

The equations of motion for $\varphi$ and ${\bf M}$ are
\begin{equation}\label{eq-varphi}
C\left(\frac{\Phi_0}{2\pi}\right)^2 \ddot{\varphi} +\frac{1}{R}\left(\frac{\Phi_0}{2\pi}\right)^2\dot{\varphi}  = -\frac{\partial U}{\partial \varphi} 
\end{equation}
\begin{equation}\label{eq-M}
\frac{d{\bf M}}{dt} = \gamma {\bf M}\times{\bf B}_{\rm eff} + \frac{\alpha}{M_0}\left({\bf M}\times\frac{d{\bf M}}{dt}\right),
\end{equation}
where $C$ and $R$ are the capacitance and the resistance of the junction, 
\begin{equation}
{\bf H}_{\rm eff} = -\frac{1}{V}\frac{\partial U}{\partial {\bf M}}
\end{equation}
is the effective field acting on the magnetic moment, and $\alpha$ is the damping parameter. To allow for quantum tunneling the junction must be of the smallest possible size, in which case its capacitance $C$ can be safely neglected. Quantum tunneling of ${\bf M}$ is carried out by the instanton solution of Eqs. (\ref{eq-varphi}) and (\ref{eq-M}) in imaginary time, $\tau = it$ at $R=0$ and $\alpha = 0$. In the resulting semiclassical equations $\varphi$ is a slave variable that follows the imaginary-time dynamics of ${\bf M}$ according to
\begin{equation}\label{eq-I}
\sin\left(\varphi - \varphi_0 \frac{M_y}{M_0}\right) = \frac{I}{I_c},
\end{equation}
making the effect of the junction on ${\bf M}$ mathematically equivalent to the effect of the external magnetic field 
\begin{equation}
{\bf B}_I = \varphi_0\frac{ E_J}{M_0} \left(\frac{I}{I_c}\right)\hat{\bf y} = \varphi_0 \frac{\Phi_0 I}{2\pi M_0}\hat{\bf y}
\end{equation}

At $I = 0$ the instanton solution of Eq.\ (\ref{eq-M}) for ${\bf M} = M_0(\sin\theta\cos\phi,\sin\theta\sin\phi,\cos\theta)$ is given by \cite{EC-PRL88,MQT-book,Lectures}
\begin{equation}
\sin\phi = \frac{\sinh(\omega_0\tau)}{\sqrt{\lambda + \cosh^2(\omega_0\tau)}}, \quad \cos\theta = \frac{\sqrt{\lambda}\cos\phi}{\sqrt{1+\lambda\sin^2\phi}}
\end{equation}
where $\omega_0  =[\omega_{\parallel}(\omega_{\parallel}+\omega_{\perp})]^{1/2}$, $\lambda = {\omega_{\parallel}}/{\omega_{\perp}}$, $\omega_{\parallel,\perp} ={2\gamma K_{\parallel,\perp}}/({M_0 V})$. The instanton switches the magnetization from ${\bf M} = -M_0\hat{\bf y}$ at $\tau = -\infty$ to ${\bf M} = M_0\hat{\bf y}$ at $\tau = +\infty$. Path integration of $\exp(i\int dt {\cal{L}}/\hbar)$ around the instanton with the Lagrangian
\begin{equation}\label{action}
{\cal{L}} = \frac{M_0V}{\gamma}\dot{\phi}(\cos\theta - 1) - U(\varphi, \theta, \phi)
\end{equation}
gives for the tunnel splitting 
\begin{equation}
\Delta_0 = Ae^{-B},  \; A \sim \hbar\omega_0,  \; B = 2S\ln(\sqrt{\lambda} +\sqrt{1+\lambda}),
\end{equation}
where $S = M_0 V/(\hbar \gamma)$ is the total spin of the ferromagnetic layer. Strong easy plane anisotropy (in which case $\lambda \ll 1$ and $B = 2\sqrt{\lambda}S$) is required to allow observation  of tunneling of a macroscopically large $S$. 

In the low energy domain the problem can be recast as a two-state problem. Projecting Eq.\ (\ref{U}) onto the two magnetic states with ${\bf M}$ along the $Y$-axis one obtains a two-state Hamiltonian  
\begin{eqnarray}\label{trancated}
H & = & -E_J\cos(\varphi - \varphi_0\sigma_y) - \frac{\Delta_0}{2} \sigma_x \nonumber \\
& = & -E_J\cos\varphi\cos\varphi_0  - {\bf b}_{\rm eff} \cdot \frac{\bm \sigma}{2}
\end{eqnarray}
with
\begin{equation}
{\bf b}_{\rm eff} = \Delta_0 \hat{\bf x} + 2E_J\sin\varphi\sin\varphi_0 \hat{\bf y}
\end{equation}
Here ${\bm \sigma}$ is a spin-$\frac{1}{2}$ operator satisfying 
\begin{equation}
\hbar \frac{d{\bm \sigma}}{dt} = i[H,{\bm \sigma}] =  {\bm \sigma} \times {\bf b}_{\rm eff}
\end{equation}
For the components of ${\bm \sigma}$ one has
\begin{eqnarray}
\hbar \frac{d\sigma_x}{dt} & = &  -
2E_J\sin\varphi\sin\varphi_0 \sigma_z \label{sigma-x}\\ 
\hbar \frac{d\sigma_y}{dt} & = &  \Delta_0 \sigma_z   \label{sigma-y}\\
\hbar \frac{d\sigma_z}{dt} & = & 2E_J\sin\varphi\sin\varphi_0 \sigma_x - \Delta_0 \sigma_y  \label{sigma-z}
\end{eqnarray}
These equations also hold for the expectation values of the components of ${\bm \sigma}$. It is easy to see that they conserve the length of ${\bm \sigma}$ ($|{\bm \sigma}| = 1$). They must be accompanied by the dynamical equation (\ref{eq-phi}) for $\varphi$ that at $C = 0$ reads
\begin{equation}\label{varphi-R}
 \frac{1}{R}\left(\frac{\Phi_0}{2\pi}\right)^2\frac{d\varphi}{dt} =  -E_J(\sin\varphi\cos\varphi_0 - \cos\varphi\sin\varphi_0 \sigma_y)
\end{equation}

In the practical limit of $|\varphi_0| \ll 1$, equations (\ref{sigma-x})-(\ref{varphi-R}), linearized near the ground state $\sigma_x = 1$, are
\begin{eqnarray}
&&\frac{1}{R}\left(\frac{\Phi_0}{2\pi}\right)\frac{d\varphi}{dt}  =  -I_c(\varphi - \varphi_0 \sigma_y) \\
&&\hbar \frac{d\sigma_y}{dt}  =   \Delta_0 \sigma_z, \quad
\hbar \frac{d\sigma_z}{dt}  =  \frac{I_c\Phi_0}{\pi}\varphi_0 \varphi - \Delta_0 \sigma_y   
\end{eqnarray}
Writing  $\varphi, \sigma_{y,z} \propto \exp(-i\omega t)$ and equating the determinant of the resulting equations to zero we get
\begin{equation}\label{determinant}
\left[\frac{-i\omega}{R}\left(\frac{\Phi_0}{2\pi}\right)+I_c\right][(\hbar\omega)^2-\Delta_0^2] + \frac{I_c^2\Phi_0}{\pi}\phi_0^2\Delta_0 = 0
\end{equation}
We look for a solution in the form
\begin{equation}\label{frequency}
\omega = \frac{\Delta_{\rm eff}}{\hbar} - i\Gamma
\end{equation}
with $\hbar \Gamma \ll \Delta_{\rm eff}$. Substituting Eq.\ (\ref{frequency}) into Eq.\ (\ref{determinant}) one obtains
\begin{equation}\label{renormalized}
\Delta_{\rm eff} =  \sqrt{\Delta_0\left(\Delta_0 - 2E_J\varphi_0^2\right)}
\end{equation}
for the tunnel splitting and
\begin{equation}\label{Gamma}
\Gamma = \left(\frac{\varphi_0^2}{\hbar R}\right)\left(\frac{\Phi_0}{2\pi}\right)^2\frac{\Delta_{\rm eff}}{\hbar}
\end{equation}
for the decoherence rate provided by the finite resistance of the junction. 

Eq.\ (\ref{renormalized}) shows that interaction of the magnetic moment with the superconducting order parameter renormalizes the tunnel splitting, $\Delta_{0} \rightarrow \Delta_{\rm eff}$. At a sufficiently large interaction, 
\begin{equation}
2E_J\varphi_0^2 > \Delta_0,
\end{equation} 
the tunneling freezes, that is, $\Delta_{\rm eff} = 0$. Smallness of $\varphi_0$ insures practicality of such a regime. 

At $2E_J\varphi_0^2 < \Delta_0$ the system prepared in a state with a definite orientation of ${\bf M}$ along the $Y$-axis begins to oscillate with the expectation values of $M_y$ and $\varphi$ satisfying
\begin{equation}
M_y = M_0e^{-\Gamma t}\cos\left(\frac{\Delta_{\rm eff}}{\hbar}t\right), \quad \varphi = \varphi_0e^{-\Gamma t}\cos\left(\frac{\Delta_{\rm eff}}{\hbar}t\right)
\end{equation}
Oscillations decay at the damping rate $\Gamma$, with the quality factor given by
\begin{equation}
Q = \left(\frac{2\pi}{\Phi_0}\right)^2\left(\frac{\hbar R}{\varphi_0^2}\right)
\end{equation} 
At $\varphi_0 \sim 0.1$ the estimate is $Q \sim 0.1R(\Omega)$, which can be quite high for an insulating ferromagnetic layer. Oscillations of $\varphi$ should generate the oscillating ac voltage across the junction,
\begin{equation}
{\cal{V}} = \frac{\hbar}{2e}\frac{d\varphi}{dt} \approx -\varphi_0\frac{ \Delta_{\rm eff}}{2e}e^{-\Gamma t}\sin\left(\frac{\Delta_{\rm eff}}{\hbar}t\right)
\end{equation}
At $\varphi_0 \sim 0.1$ and $\Delta_{\rm eff} \sim 0.1$K the initial ($t = 0$) amplitude of the ac voltage would be in the microvolt range, while the frequency, $\Delta{\rm eff}/(2\pi \hbar)$, would be in the GHz range, which should not be difficult to detect. 

To illustrate the possibility to control the magnetic tunneling by the superconducting current consider the case of a uniaxial magnetic anisotropy with an easy axis along the $Z$-direction, 
\begin{equation}
U_M =  - \frac{1}{2}K_{\parallel}V\left(\frac{M_z}{M_0}\right)^2
\end{equation}
studied in Ref.\ \onlinecite{Buzdin-PRL09}. With account of Eq.\ (\ref{eq-I}) one obtains the following dynamical equations for the spherical angles describing the orientation of ${\bf M}$ in space
\begin{eqnarray}
\frac{d\phi}{dt'} & = & -\left( \sin\theta - \frac{I}{I_0}\sin\phi \right)\frac{\cos\theta}{\sin\theta} \label{eq-phi}\\
\frac{d\theta}{dt'} & = & -\frac{I}{I_0} \cos\phi \label{eq-theta},
\end{eqnarray}
where $t'=\omega_{\rm FMR}t$ with $\omega_{\rm FMR} =\gamma K_{\parallel}/M_0$ being the frequency of the ferriomagnetic resonance at $I = 0$, and
\begin{equation}\label{I-0}
I_0 = \frac{K_{\parallel}V}{\varphi_0 E_J} I_c
\end{equation}
Note that depending on the values of parameters entering Eq.\ (\ref{I-0}) $I_0$ can be smaller or greater than $I_c$. At $I = 0$ the degenerate ground state corresponds to the magnetic moment parallel ($\theta = 0$) or antiparallel ($\theta = \pi$) to the $Z$-axis. At $I < I_0$ (assuming also that $I < I_c$) the degenerate ground states are
\begin{equation}\label{degenerate}
\phi = \frac{\pi}{2}, \quad \sin\theta = \frac{I}{I_0}, \quad \cos\theta = \pm\sqrt{1-\left(\frac{I}{I_0}\right)^2}
\end{equation}
At $I_0 < I < I_c$ the non-degenerate ground state is $\phi = \pi/2, \theta = \pi/2$, corresponding to the magnetic moment looking in the $Y$-direction.  

In the case of $I_0 < I_c$, when $I$ is close but smaller than $I_0$, the energy barrier between the degenerate states in Eq.\ (\ref{degenerate}) is small,
\begin{equation}
U(I) = \frac{1}{2}K_{\parallel}V\epsilon^2, \qquad \epsilon = 1-\frac{I}{I_0} \ll 1
\end{equation}
We are interested in the quantum tunneling between degenerate classical ground states: $\phi = \pi/2, \theta = \pi/2 \pm \sqrt{2\epsilon}$. Substituting $\theta = \pi/2 + \beta$ and $\phi = \pi/2 + \alpha$, with $|\alpha|, |\beta| \ll 1$, into Eqs.\ (\ref{eq-phi}) and (\ref{eq-theta}) one obtains
\begin{equation}
\frac{d\alpha}{dt'}  =  \left(\epsilon - \frac{\beta^2}{2}\right), \quad \frac{d\beta}{dt'} =  \alpha
\end{equation}
Here we have taken into account (see below) that $\alpha \sim \epsilon$ while $\beta \sim \sqrt{\epsilon}$, making $\alpha \ll \beta$. 
Combining the two equations, introducing $\bar{\beta} = \beta/\sqrt{2\epsilon}$ and the imaginary time $\bar{\tau} = it'\sqrt{\epsilon/2}$ we get 
\begin{equation}
\frac{d\bar{\beta}}{d\bar{\tau}} = 1-\bar{\beta}^2
\end{equation}
which has the instanton solution $\bar{\beta} = \tanh \bar{\tau}$,
\begin{equation}
\beta(\tau) = \sqrt{2\epsilon}\tanh\left(\sqrt{\frac{\epsilon}{2}}\omega_{\rm FMR}\tau\right)
\end{equation}
that switches $\beta$ between $-\sqrt{2\epsilon}$ at $\tau = -\infty$ to $\sqrt{2\epsilon}$ at $\tau = \infty$. 

Path integration of $\exp(i\int dt {\cal{L}}/\hbar)$ around the instanton with the Lagrangian given by Eq.\ (\ref{action}) yields for the tunnel spliting $\Delta = Ae^{-B}$ with
\begin{equation}
A \sim  \hbar\omega_{\rm FMR}\left(1-\frac{I}{I_0}\right)^{1/2},  \; B = \frac{4\sqrt{2}}{3}S\left(1-\frac{I}{I_0}\right)^{3/2}
\end{equation}
Notice that the tunneling rate in this case depends exponentially on the superconducting current. However, contrary to the previously studied case of biaxial magnetic anisotropy, quantum oscillations of the magnetic moment between classically degenerate states $\phi = \pi/2, \theta = \pi/2 \pm \sqrt{2\epsilon}$ formed by the uniaxial anisotropy along the $Z$-axis and a superconducting current in the $X$ direction (see Fig.\ \ref{geometry}) do not produce oscillations of $\varphi$. Consequently, they do not generate any voltage across the junction. One way to detect such a tunneling would be with the help of a SQUID  loop sensitive to a small $Z$-component of the magnetic moment that changes sign in the tunneling event.  Nano-SQUIDs permit detection of the change in the magnetic moment of only a few Bohr magnetons \cite{Wernsdorfer}. In our case the change would be much greater. In experiments with single-domain magnetic nanoparticles and molecules, the energy barrier and the tunneling rate were controlled by a strong magnetic field, which negatively affected the performance of the SQUID. The advantage of the $\varphi_0$-junction is that the barrier and the tunneling rate can be accurately controlled by the superconducting current, with no need for the external magnetic field. 

This work has been supported by the grant No. DE-FG02-93ER45487 funded by the U.S. Department of Energy, Office of Science.

\end{document}